\documentclass[twocolumn,showpacs,preprintnumbers,prb]{revtex4}%

\usepackage{graphics}
\usepackage{longtable}
\usepackage{graphicx}
\usepackage{dcolumn}
\usepackage{bm}
\usepackage{amsmath}
\usepackage{amsfonts}
\usepackage{amssymb}

\begin{document}

\title{Local Behavior of the First-Order Gradient Correction to the
Thomas-Fermi Kinetic Energy Functional}
\author{David Garc\'{\i}a-Aldea}
\email{dgaldea@fisfun.uned.es}
\affiliation{Departamento de F\'{\i}sica Fundamental. UNED. Apartado 60.141. 
E-28080 Madrid (Spain). }
\author{T. Mart\'{\i}n-Blas}
\email{teresa.martin@upm.es}
\affiliation{Departamento de Ciencias B\'{a}sicas, E.U.I.T. Forestal, 
Universidad Polit\'{e}cnica de Madrid. 
Ciudad Universitaria. E-28040 MADRID (Spain). }
\author{J. E. Alvarellos}
\email{jealvar@fisfun.uned.es}
\affiliation{Departamento de F\'{\i}sica Fundamental. UNED. Apartado 60.141. 
E-28080 Madrid (Spain). }
\date{\today }

\begin{abstract}
The first order gradient correction to the Thomas-Fermi functional, 
proposed
by Haq, Chattaraj and Deb (Chem. Phys. Lett. \textbf{81}, 8031, 1984) 
has been studied by evaluating both the total kinetic energy and 
the local kinetic energy density. 
For testing the kinetic energy density 
we evaluate its deviation from an exact result 
through a \emph{quality factor}, a
parameter that reflects the quality of the functionals in a better way than
their relative errors. 
The study is performed on two different systems:
light atoms (up to $Z=18$) and a noninteracting model of fermions confined
in a Coulombic-type potential. 
It is found than this approximation gives very low relative errors and a 
better local behavior than any of the usual 
generalized gradient approximation semilocal 
kinetic density functionals.
\end{abstract}

\maketitle




\section{Introduction}

\label{Sec:Intro}

Density Functional Theory, originally developed by 
Hohenberg and Kohn (HK), \cite{1964HK} 
replaces the main role of the wave function in the 
description of many electron systems and proves that the 
particle density 
$n\left( \mathbf{r}\right) $ 
determines all the properties of the system through the functional 
of the total energy $E\left[ n\right] $. 
When $E\left[n\right] $ is minimized, the density and the energy of 
the ground state can be found. 
The functional for the total energy of the interacting system is
unknown, but Kohn and Sham (KS)\cite{1965KS} devised a method for 
the minimization of $E\left[ n\right] $ that uses the concept of 
noninteracting KS orbitals 
$\varphi _{i}\left( \mathbf{r}\right) $ that yield the same
density than the interacting system. 
The KS scheme allows to divide $E\left[n\right] $ in four terms, 
and one of them is the kinetic energy density functional (KEDF) 
of the KS noninteracting system, $T_{S}\left[ n\right] $, 
that can be exactly evaluated in terms of the KS orbitals. 

In order to provide insight into the physics of many-body systems, 
we are interested in elucidating the formal properties of the KEDF. 
As explained below, the Thomas-Fermi (TF) functional is a local 
density approximation that has been corrected with other terms 
that usually depend on the variations of the density, normally 
including second-order gradient terms 
(see Ref.~\onlinecite{2007DGAJEA1} for a general review on the results 
obtained with a number of semilocal kinetic functionals).
In this paper we will study the kinetic
energy functional that includes the first-order gradient correction to the
TF functional proposed by Haq, Chattaraj and Deb.%
\cite{1984HCD, 1988DC, 1990Chattaraj, 1992DC, 1992ZCPL}
To assess the quality of this first-order gradient correction we
will evaluate both the total kinetic energy and the local behavior 
of the approximate kinetic energy density (KED). 
The study of the KED 
(a function that depends on the spatial coordinates) 
is made through a 
\emph{quality factor} $\sigma $, a parameter
that reflects the quality of the functionals in a better way than 
their relative errors.%
\cite{2007DGAJEA1, 2007DGAJEA2, 2008DGAJEA1, 2008DGAJEA2, 2009TMBDGAJEA} 
It may be expected that the functional which systematically gives
the better KED would probably also yield another properties more 
accurately.

In order to make the study of how this first-order gradient functional 
works for localized systems we test two simple systems: 
a noninteracting model (fermions confined in a Coulombic-type 
potential) 
and an interacting electron system (light atoms, from He to Ar). 

\subsection{Kinetic Energy Functionals}

As commented, the KS scheme allows to define the KEDF of the KS
noninteracting system, $T_{S}\left[ n\right] $, for any many particle
system. 
But, following the original HK idea, 
many attempts to write the kinetic energy as a functional, using
exclusively the electron density, have been made. The very first attempt 
was a statistical method 
-- the Thomas-Fermi (TF) functional, proposed by Thomas%
\cite{1927Th} and Fermi\cite{1927Fer} -- that has the form:%
\begin{equation}
T_{TF}\left[ n\right] =C_{TF}\int {d\mathbf{r}n(\mathbf{r})^{5/3}\mathbf{,}} 
\quad \quad 
C_{TF}=\frac{3}{10}(3\pi ^{2})^{2/3}.
\end{equation}%
TF is exact for a free electron gas or an uniform system, but this local
density approximation yields usually big relative errors --around 10\%-- for
systems such as atoms and molecules. 

Another well-known functional, due to von Weizs\"{a}cker (vW)\cite%
{1935Weiz}, is exact for any system described by a single orbital, 
\begin{equation}
T_{vW}[n]=\frac{1}{8}\int {d\mathbf{r}\frac{|\mathbf{\nabla }n(\mathbf{r}%
)|^{2}}{n(\mathbf{r})}}.
\end{equation}

The TF functional is usually corrected by functionals that depend on the
variations of the density, as in the \emph{second-order Gradient Expansion
Approximation})\cite{1957Kirzhnits} 
(GEA2, where the vW term is weighted by $1/9$) 
\begin{equation}
T_{GEA2}\left[ n \right] = 
C_{TF} \int d\mathbf{r} n(\mathbf{r})^{5/3} 
+ \frac{1}{72} \int {d\mathbf{r} \frac{|\mathbf{\nabla }%
n(\mathbf{r})|^{2}}{n(\mathbf{r})},}
\end{equation}
that usually yields small relative errors --usually lower than 1\%-- 
when applied to localized systems using \emph{good} densities 
(i.\ e., those obtained with accurate methods as the Hartree-Fock or 
the KS ones), 
although it gives a wrong local behavior,
misplacing the KED about $18\%$. \cite{2007DGAJEA1} 
Another functional that has obtained good results for these type of systems 
(when atomic systems and homonuclear molecules are variationally solved)%
\cite{1965YoneiTomish, 1966TomishYonei}
is the TF$\frac{1}{5}$vW one, 
\begin{eqnarray}
T_{TF\frac{1}{5}vW}\left[ n\right] 
&=& 
T_{TF}\left[ n\right] + \frac{1}{5} T_{vW}\left[ n\right] 
\notag
\\
&=& 
C_{TF}\int d\mathbf{r~}n(\mathbf{r})^{5/3} 
+ \frac{1}{40}
\int {d\mathbf{r}
\frac{|\mathbf{\nabla }n(\mathbf{r})|^{2}}{n(\mathbf{r})}}.
\end{eqnarray}
But if we use these \emph{good} electron densities,  
TF$\frac{1}{5}$vW usually yields bigger relative errors 
than GEA2 and usually
overestimates the total kinetic energy by more than $6\%$.

Beyond the GEA2 and TF$\frac{1}{5}$vW, 
a number of other semilocal functionals within the \emph{%
Generalized Gradient Approximation} (GGA) can be proposed using the
general form 
\begin{equation}
T_{GGA}\left[ n\right] =C_{TF}\int {d\mathbf{r~}n(\mathbf{r})^{5/3}~F}\left[
s\left( \mathbf{r}\right) \right].
\end{equation}%
where ${F}\left[ s\left( \mathbf{r}\right) \right] $ is an \emph{enhancement
factor} that corrects the TF functional via the adimensional reduced
gradient, 
\begin{equation*}
s\left( \mathbf{r}\right) =\frac{|\mathbf{\nabla }n(\mathbf{r})|}{n(\mathbf{r%
})^{4/3}},
\end{equation*}%
that takes into account the inhomogeneities of the system. The GGA
functionals include both the TF functional 
(with ${F}\left[ s\left( \mathbf{r}\right) \right] =1$)\ 
and the GEA2 and the TF$\frac{1}{5}$vW ones, as well as many other
semilocal functionals. 
The GGA functionals generally depend on $s\left(\mathbf{r}\right) ^{2}$, 
and their mathematical form and their results are 
cleary related to the GEA2; only some of the GGAs functionals 
yield smaller relative errors than the GEA2 but none of them 
give better KED than the TF functional. 
\cite{2007DGAJEA1} 

It is also important to know that, due to symmetry considerations in general
systems, these functionals depend only on the absolute value of the gradient
of the density (or its squared value) and never take into account the
directionality of the density variations.

\subsection{First-Order Gradient Correction}

\label{FO}

For localized systems a different functional has been proposed by 
Haq, Chattaraj and Deb (HDC)\cite{1984HCD}. 
This functional is an interesting exception, because it depends 
on the gradient of the electron density and not on its squared gradient, 
and also depends on the position vector $\mathbf{r}$ 
(differently to the other GGAs). 

The form of the proposed functional is:%
\begin{eqnarray}
{T}_{HCD-1}\left[ n\left( \mathbf{r}\right) \right] 
& = & 
C_{TF}\int d\mathbf{r~}n(\mathbf{r})^{5/3} 
- \frac{1}{40}\int d\mathbf{r} 
\frac{\mathbf{r\nabla }n(\mathbf{r})}{r^{2}} 
\notag
\\
& = & 
T_{TF}\left[ n\left( \mathbf{r}\right) \right] 
+ T_{r-1}\left[ n\left( \mathbf{r}\right) \right] ,
\end{eqnarray}%
i. e., the sum of the TF functional and an $r$-dependent correction 
${T}_{r-1}\left[ n\left( \mathbf{r}\right) \right] $.
Note that the $1/40$ weight resembles the same prefactor 
that appeared in the  TF$\frac{1}{5}$vW functional 
and that the gradient of the electron density is included in the 
$r$-dependent term, so this correction is
usually labeled as \emph{first-order gradient correction} 
(instead of the second-order corrections previously presented).

Making use of the relation%
\begin{equation*}
\int {d\mathbf{r}\frac{\mathbf{r\nabla }n(\mathbf{r})}{r^{2}}\mathbf{=-}}%
\int {d\mathbf{r}\frac{n(\mathbf{r})}{r^{2}},}
\end{equation*}%
a different first-order term is obtained, and a corresponding functional can 
be proposed, 
\begin{equation*}
{T}_{r-2}\left[ n\left( \mathbf{r}\right) \right] =
\frac{1}{40}\int {d\mathbf{r}\frac{n(\mathbf{r})}{r^{2}}.}
\end{equation*}
The functional 
$
{T}_{HCD-2}\left[ n\left( \mathbf{r}\right) \right] = 
T_{TF}\left[ n\left( \mathbf{r}\right) \right] + 
T_{r-2}\left[ n\left( \mathbf{r}\right) \right]
$ 
yields the same total kinetic energy but has a different local 
behavior.

It is also known that 
${T}_{r-1}\left[ n\left( \mathbf{r}\right) \right] $\
without the $1/40$ prefactor is a lower bound \cite{1982GadrePathak}
to the \emph{Fisher information function}, %
\cite{1925Fisher,1980SearsParrDinur,2007L} 
$I_{F}\left[ n\left( \mathbf{r}\right) \right]$, 
\begin{equation}
-\int {d\mathbf{r}\frac{\mathbf{r\nabla }%
n(\mathbf{r})}{r^{2}}}\leq \int {d \mathbf{r}\frac{|\mathbf{\nabla }%
n(\mathbf{r})|^{2}}{n(\mathbf{r})}}
= I_{F}\left[ n\left( \mathbf{r}\right) \right] .
\label{TvWbound}
\end{equation}%
On the other hand, the Fisher information function is directly related
related to the vW functional, 
\begin{equation}
T_{vW}[n]=\frac{1}{8}I_{F}\left[ n\left( \mathbf{r}\right) \right] ,
\end{equation}
so 
$\frac{1}{5}T_{vW}[n]$ is an upper bound to 
${T}_{r-1}\left[ n\left( \mathbf{r}\right) \right] $
and 
${T}_{TF\frac{1}{5}vW}\left[n\left( \mathbf{r}\right) \right] $ 
becomes an upper bound to the HCD functional, 
\begin{equation*}
{T}_{HCD}\left[ n\left( \mathbf{r}\right) \right] \leq {T}_{TF\frac{1}{5}vW}%
\left[ n\left( \mathbf{r}\right) \right].
\end{equation*}%
For all these results, the functional 
${T}_{HCD}\left[ n\left( \mathbf{r}\right) \right] $
could be a promising functional yielding small relative errors. 

Note that ${T}_{r-2}\left[ n\left( \mathbf{r}\right) \right] $ 
is the average of 
$r^{-2}$ over the system, 
$\left\langle r^{-2}\right\rangle =\int %
{d\mathbf{r}\frac{n(\mathbf{r})}{r^{2}}}$, 
with a $1/40$ prefactor. 
This quantity is related to exact bounds for the 
von Weizs\"{a}cker functional, 
being particularly interesting the following one:%
\begin{equation}
\frac{1}{8}\left\langle r^{-2}\right\rangle <T_{vW}\left[ n\left( \mathbf{r}%
\right) \right] <\frac{1}{4}\left\langle r^{-2}\right\rangle .
\end{equation}
The lower bound is of variational origin while the upper bound is 
based on the complete monotonicity hypothesis.
(see a general discussion on this topic in 
Ref.~(\onlinecite{1994RomeraDehesa}) and references therein).

We must note that the choice of $1/40$ as a prefactor has 
no theoretical justification and was chosen by trial and 
error. 
But it is possible to make a qualitative exercise to make a guess 
of the approximate value of this prefactor. 
We can argue using the previous
inequality that the kinetic energy obtained with the vW functional 
should be close to the mean of the two bounds, i.e. 
\begin{equation}
\frac{3}{16}\left\langle r^{-2}\right\rangle \approx %
T_{vW}\left[ n\left( \mathbf{r}\right) \right] .
\end{equation}
On the other hand, it is well known that the TF functional can be 
quite accurately corrected by adding a weighted vW term  
(e.g. the GEA2 functional usually yields errors lower than 1\%). 
And by replacing the vW functional by 
$\frac{3}{16}\left\langle r^{-2}\right\rangle $, 
we add to the TF functional a correction that is 
$\frac{1}{48} \left\langle r^{-2}\right\rangle$. 
This value $1/48$ is quite close to the proposed $1/40$ 
(a difference of 17\%, for a correction that represents about  
8\% of the total kinetic energy).

\subsection{The Relative Errors and the Local Kinetic Behavior}

\label{Sec:study}

As pointed out before, the present paper is devoted to the study of 
the two functionals proposed in Sect.~(\ref{FO}), 
${T}_{HCD}\left[n\left( \mathbf{r}\right) \right] $, 
with an analysis of both their total
kinetic energy and their KED.
We study two different systems with spherical symmetric densities.
Firstly, we use the electron densities of light atoms from 
$Z=2$ to $Z=18$.
As a second system, we will study the noninteracting fermion 
system confined in a Coulomb-type potential presented in 
Ref.~(\onlinecite{2009TMBDGAJEA}). 
In both cases, we have worked with the analytical expressions of the
single-particle orbitals in order to evaluate not only the relative 
errors obtained by the approximate KEDFs, 
but also compare their KED to the exact KS results (that can be 
directly obtained from these orbitals). 
To make that comparison we have introduced 
\cite{2007DGAJEA1, 2007DGAJEA2, 2008DGAJEA1, 2008DGAJEA2, 2009TMBDGAJEA} 
a quantity $\sigma$ that takes into account the local differences, 
focuses on the error cancellations, and reflects the fact 
that the KED is not univocally defined.

Let's do a brief review of the method. 
For a given set of $N$ orbitals $\varphi
_{i}(\mathbf{r})$ we can use the usual orbital-based KED, 
$t_{S}^{I}\left( \mathbf{r}\right) =%
\frac{1}{2}\sum_{i=1}^{N}%
\left\vert \nabla \varphi _{i}(\mathbf{r})\right\vert ^{2}$, 
or another common definition of the KED, 
$t_{S}^{II}\left( \mathbf{r}\right) =-\frac{1}{2}%
\sum_{i=1}^{N}\varphi _{i}^{\ast }(\mathbf{r})%
\nabla ^{2}\varphi _{i}(\mathbf{r})$. 
But for this set of $N$ orbitals we can construct an infinite
number of valid KEDs using a linear combination of 
$t_{S}^{I}$ and $t_{S}^{II}$, 
\begin{eqnarray}
t_{S}^{L,\alpha }(\mathbf{r}) &=&%
\left( 1-\alpha \right) t_{S}^{I}\left(\mathbf{r}\right) 
+ \alpha t_{S}^{II}\left( \mathbf{r}\right)%
\notag \\
&=&%
\left( 1-\alpha \right) \frac{1}{2}\sum_{i=1}^{N}%
\left\vert \nabla\varphi _{i}(\mathbf{r})\right\vert ^{2}
\\
& & 
+ \alpha \sum_{i=1}^{N}\left( -\frac{1}{2}%
\varphi _{i}^{\ast }\left( \mathbf{r}\right) \nabla ^{2}%
\varphi_{i}\left( \mathbf{r}\right) \right)%
\notag \\
&=&%
t_{S}^{I}\left( \mathbf{r}\right) -%
\frac{1}{4}\alpha \nabla ^{2}n\left(\mathbf{r}\right) %
\label{Eq:tslap}
\end{eqnarray}%
(all of them are valid KED, see Ref.~(\onlinecite{1996YLW})).
So, $t_{S}^{L,\alpha }(\mathbf{r})$ 
has a first term 
$t_{S}^{I}\left( \mathbf{r}\right) $ 
given by the orbitals plus a term proportional to the 
laplacian of the electron density (the laplacian
integrates to zero in the whole space). 
The real parameter $\alpha $ can be arbitrarily varied and the 
family of KEDs described by 
$t_{S}^{L,\alpha }(\mathbf{r})$ 
includes 
$t_{S}^{I}\left( \mathbf{r}\right) $ 
(when $\alpha =0$) 
and $t_{S}^{II}\left( \mathbf{r}\right) $ 
(for $\alpha =1$). 

When an approximate KEFD is used,each value of 
$\alpha $ defines a \textsl{quality factor} 
\begin{equation}
\sigma \left( \alpha \right) =%
\frac{\int d\mathbf{r}\left\vert
t_{S}^{L,\alpha }(\mathbf{r})-t_{S}^{func}(\mathbf{r})\right\vert }%
{T_{S}[n]},%
\label{Eq:sigma}
\end{equation}%
being $t_{S}^{func}(\mathbf{r})$ the approximate KED of the kinetic
functional. 
$\sigma $ represents a summation of the absolute value of
the differences between KEDs, so it measures how the approximated 
KED is misplaced with respect
to the distribution $t_{S}^{L,\alpha }(\mathbf{r})$. 
Note that $\sigma $ is zero if the approximate and exact
distributions of KED are identical, 
and when the larger the value of $\sigma $ is 
the bigger the differences between the two are. 
For each approximate functional we will obtain a 
set of values of $\sigma \left(\alpha \right)$ 
and in order to compare the KEDs we choose the best value of 
$\alpha $\ for each functional, i. e. the value of 
$\alpha _{min}$ that yields the minimum value
of $\sigma \left( \alpha \right) $; 
the KED has its closest $t_{S}^{L}(\mathbf{r})$
by evaluating Eq.~(\ref{Eq:tslap}) for this value $\alpha _{min}$.
For details on this measure $\sigma $, the behavior of GGA
functionals and the non uniqueness in the definition of the exact KED, see
Ref.~(\onlinecite{2007DGAJEA1}) and references therein.

\section{Results and Discussion}

\label{Sec:results}

In order to assess the quality of the two HCD, 
$r$-dependent functionals,  
introduced in Sec.~\ref{FO} (HCD-1 and HCD-2), 
we are going to compare
both the relative errors in the kinetic energies and the quality 
factor $\sigma $ of the KED obtained with these functionals 
to the values of the same magnitudes obtained with the TF, GEA2 and 
TF$\frac{1}{5}$vW functionals.

We have applied the functionals to two different systems. 
Firstly we study the atoms from He to Ar (from $Z=2$ to $Z=18)$. 
We will approximate the atomic orbitals by orthogonalized 
Slater orbitals, because they do not add any spurious 
oscillations in the laplacian of the density 
and both the correct cusp behavior and the density
decay for $r\rightarrow \infty $ can be achieved 
(the Gaussian-type basis sets have well known deficiencies, 
they cannot reproduce the cusps of electronic density 
and electronic orbitals on nuclei and cannot provide the correct 
asymptotic behavior at long distances from the nuclei).
Being the orbitals known, the local behavior of the KED can be exactly
studied. 
The values of the exponents for the Slater orbitals
\cite{b1997Atkins,1963CR} 
and the corresponding exact total kinetic energy
for the atoms used in this paper are shown in 
Table~\ref{Table:Zeff}.

\begin{table*}[tbp] 
\centering

\caption{Values of the Slater orbital exponents and 
the corresponding exact total kinetic energy for 
the atoms studied in the paper.}%
\label{Table:Zeff}%

\begin{tabular}{c|ccccccccc}
\hline
\hline
      Atom &         He &         Li &         Be &          B &          C &   %
         N &          O &          F &         Ne \\
\hline
        1s &     1.6875 &     2.6906 &     3.6848 &     4.6795 &     5.6727 &   %
    6.6651 &     7.6579 &     8.6501 &     9.6421 \\

        2s &            &     0.6396 &     0.9560 &     1.2881 &     1.6083 &   %
    1.9237 &     2.2458 &     2.5638 &     2.8792 \\

        2p &            &            &            &     1.2107 &     1.5679 &   %
    1.9170 &     2.2266 &     2.5500 &     2.8792 \\

\hline

  Exact KE &     2.8477 &     7.4421 &     14.556 &     24.498 &     37.620 &   %
    54.281 &     74.538 &     98.996 &     127.80 \\
\hline
\hline
      Atom &            &         Na &         Mg &         Al &         Si &   %
         P &          S &         Cl &         Ar \\
\hline
        1s &            &    10.6259 &    11.6089 &     12.591 &    13.5745 &   %
   14.5578 &    15.5409 &    16.5239 &    17.5075 \\

        2s &            &     3.2857 &     3.6960 &     4.1068 &     4.5100 &   %
    4.9125 &     5.3144 &     5.7152 &     6.1152 \\

        2p &            &     3.4009 &     3.9129 &     4.4817 &     4.9725 &   %
    5.4806 &     5.9885 &     6.4966 &     7.0041 \\

        3s &            &     0.8358 &     1.1025 &     1.3724 &     1.6344 &   %
    1.8806 &     2.1223 &     2.3561 &     2.5856 \\

        3p &            &            &            &     1.3552 &     1.4284 &   %
    1.6288 &     1.8273 &     2.0387 &     2.2547 \\

\hline

  Exact KE &            &     161.14 &     198.82 &     242.86 &     289,19 &   %
    340.88 &     397.34 &     458.88 &     525.59 \\
\hline
\hline

\end{tabular}  

\end{table*}

\begin{figure}[tbp]
\begin{center}
\includegraphics[height=3.1367in,width=3.6in]{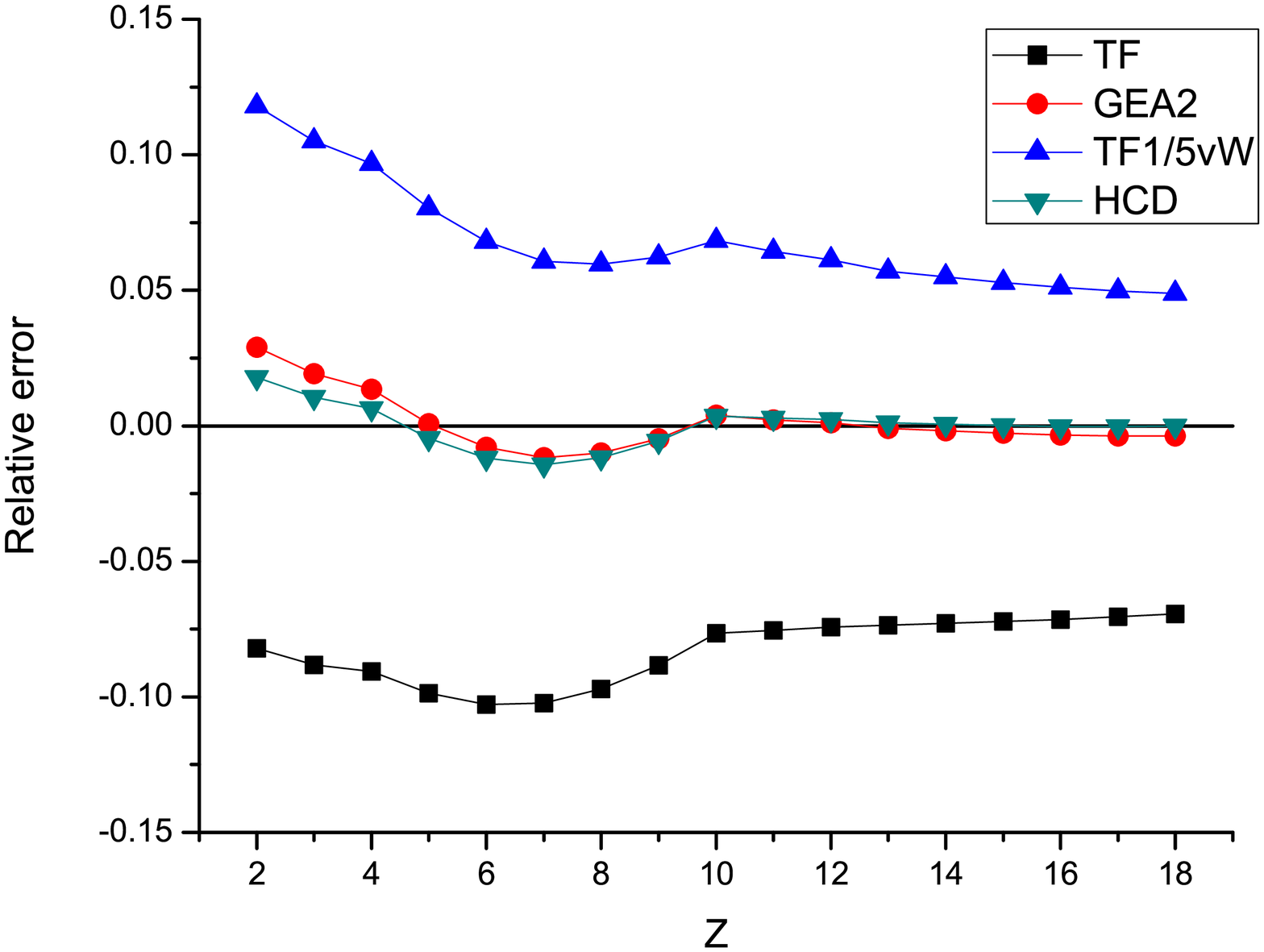}
\end{center}
\caption{ (Color online) 
Relative errors of the TF, GEA2, TF$\frac{1}{5}$vW and the 
first order correction HCD functionals 
for atoms from He to Ar.
}
\label{Fig:err_atoms}
\end{figure}

Fig.~\ref{Fig:err_atoms} shows the values of the relative errors 
of the total kinetic energy yielded by the functionals for the 
atoms, as a function of the atomic number $Z$. 
The best results are obtained for the GEA2 and the two HCD 
functionals, with quite close errors. 
The TF functional gives a negative relative error, whereas the 
TF$\frac{1}{5}$vW shows a similar trend, but with positive errors 
and smaller absolute values. 
In all cases, the relative errors for all theses KEDFs are almost 
constant for $Z \geq 10$.

\begin{figure}[tbp]
\begin{center}
\includegraphics[height=3.1176in,width=3.6in]{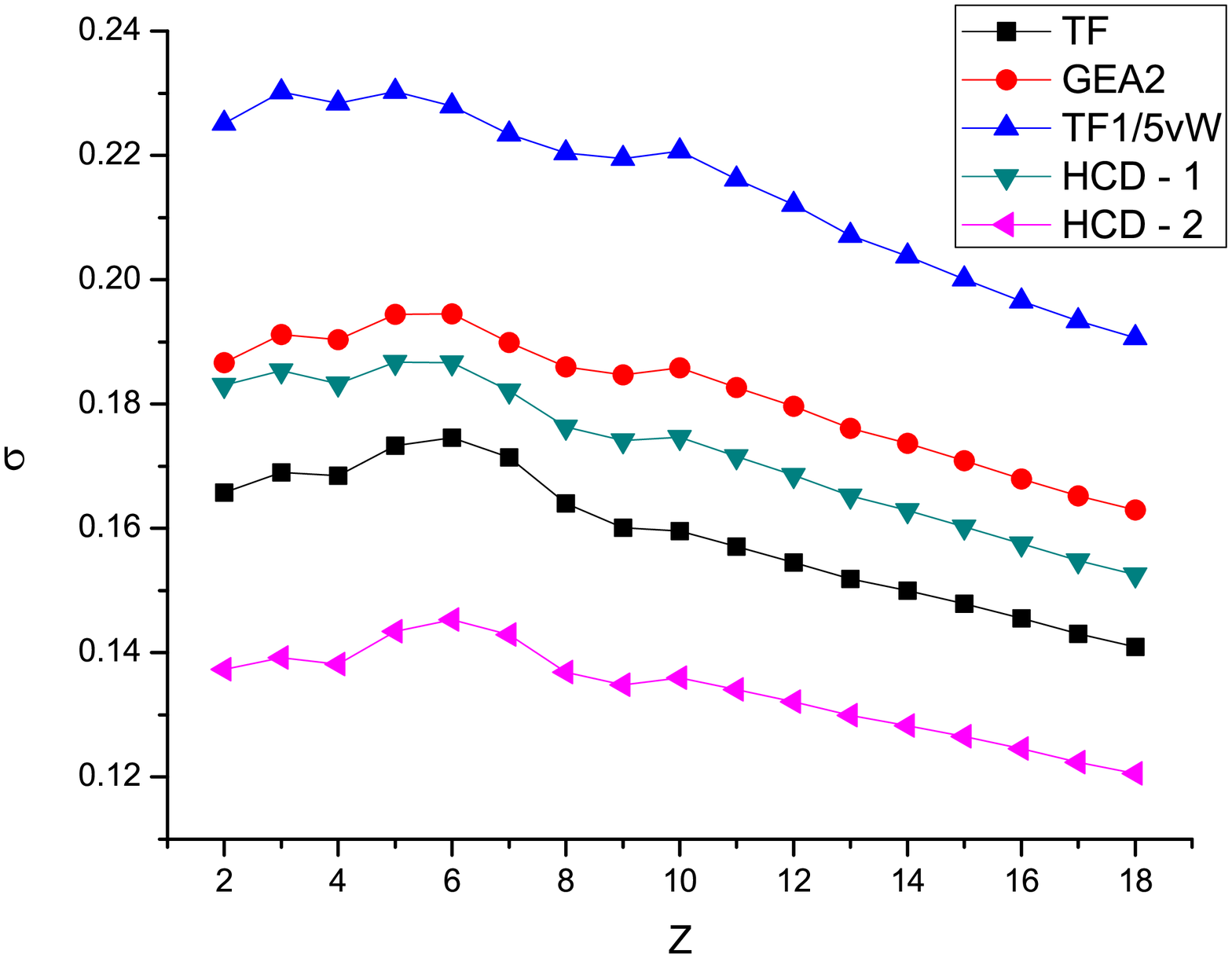}
\end{center}
\caption{ (Color online)
Quality factors $\sigma$ obtained with the TF, GEA2, TF$\frac{1}{5}$vW 
and the first order correction HCD functionals 
for atoms from He to Ar. 
}
\label{Fig:sigma_atoms}
\end{figure}

For the atoms, the value of $\sigma$ is smaller the bigger $Z$
is, as shown in Fig.~\ref{Fig:sigma_atoms}, where the values 
of the quality factor are plotted as a function of $Z$. 
The results clearly show that the HCD-2 functional give the smallest  
values of $\sigma$, reflecting the better local behavior of this 
semilocal functional. 
We have found that the GGA corrections to the TF functional 
improve the TF results for the total kinetic energies, 
but they always give worse local KEDs, 
an unexpected failure that shows that GGA functionals are 
unable to improve the local pathologies of the TF functional. %
\cite{2007DGAJEA1} 
This failure is overcome by the HCD-2 form of the first-order 
correction, 
whereas the HCD-1 form improves the total kinetic energies, 
but not the value of $\sigma$, as can be seen in 
Figs.~\ref{Fig:err_atoms}  and \ref{Fig:sigma_atoms}.

The second test system is constructed with a set of noninteracting 
fermions that are confined in a Coulombic potential. %
\cite{2009TMBDGAJEA} 
In this case, we can also use analytical single-particle 
orbitals, those of a Coulombic-type potential. 
We will use spherically symmetric densities for this 
$N$-fermion system built up with mixed states 
and we will describe the degenerate $N$-fermion states 
that arise from the distinct values of the occupancy of 
the $l$ single-fermion states. 
These degenerate $N$-fermion states have  
a different orbital angular momentum for each configuration, 
with a corresponding different density 
$n\left( \mathbf{r}\right) $. \cite{2009TMBDGAJEA} 
As the HK theorems hold not only for densities 
that correspond to a ground-state density for some 
external potential\cite{1964HK}
but also for a weighted average of ground-state electron 
densities corresponding
to that potential, \cite{1983Lieb, 1982Levy, 1983EE, 2000UK} 
we can directly discuss the densities corresponding to a 
degenerate ground-state.%
\cite{1998Nagy, 2000UK, 2000YZA, 2005NLB}

\begin{table*}[tbp] 
\centering%

\caption{
Relative errors $\epsilon$ and optimized values of the 
quality factor $\sigma_{min}$ for all the configurations 
of the noninteracting $N$-fermion system from 
$N=1$ to $10$. 
The occupation numbers $p$ of each orbital is shown, 
as well as the values of 
$\sum_{i=1}^{N} l_{i}(l_{i}+1)$.
}%
\label{Table:TableES}%

\begin{tabular}{c|cccc|cc|cc|cc|cc|cc}

\hline
\hline
 & \multicolumn{ 4}{|c}{Orbital occupation} 
 & \multicolumn{ 2}{|c}{TF} 
 & \multicolumn{ 2}{|c}{TF$\frac{1}{5}$vW} 
 & \multicolumn{ 2}{|c}{GEA2} 
 & \multicolumn{ 2}{|c}{HCD-1} 
 & \multicolumn{ 2}{|c}{HCD-2} \\

\hline 

$N$ 
&     $p_{1s}$ 
&     $p_{2s}$ 
&     $p_{2p}$ 
&     $\sum_{i=1}^{N} l_{i}(l_{i}+1)$ 
& $\sigma_{min}$ & $\epsilon$  
& $\sigma_{min}$ & $\epsilon$
& $\sigma_{min}$ & $\epsilon$
& $\sigma_{min}$ & $\epsilon$
& $\sigma_{min}$ & $\epsilon$
\\

\hline

 1 &   1 &  0 &   0 &   0 &    0.422  &  -0.422 &    0.222 &  -0.222 &    0.311  &   -0.311  &  0.322  & -0.322 &  0.330 & -0.322 \\
\hline                                                                              
                                                                                    
 2 &   2 &  0 &   0 &   0 &    0.166  &  -0.082 &    0.225 &   0.118 &    0.187  &    0.029  &  0.171  &  0.018 &  0.137 &  0.018 \\
\hline                                                                              
                                                                                    
 3 &   2 &  1 &   0 &   0 &    0.179  &  -0.107 &    0.236 &   0.073 &    0.200  &   -0.007  &  0.182  & -0.013 &  0.148 & -0.013 \\
                                                                                    
 3 &   2 &  0 &   1 &   2 &    0.198  &  -0.126 &    0.244 &   0.041 &    0.215  &   -0.033  &  0.201  & -0.036 &  0.l70 & -0.036 \\
\hline                                                                              
                                                                                    
 4 &   2 &  2 &   0 &   0 &    0.180  &  -0.114 &    0.233 &   0.057 &    0.199  &   -0.019  &  0.182  & -0.024 &  0.149 & -0.024 \\
                                                                                    
 4 &   2 &  1 &   1 &   2 &    0.196  &  -0.132 &    0.238 &   0.026 &    0.211  &   -0.044  &  0.197  & -0.046 &  0.166 & -0.046 \\
                                                                                    
 4 &   2 &  0 &   2 &   4 &    0.209  &  -0.145 &    0.241 &   0.003 &    0.220  &   -0.062  &  0.210  & -0.061 &  0.182 & -0.061 \\
\hline                                                                              
                                                                                    
 5 &   2 &  2 &   1 &   2 &    0.187  &  -0.127 &    0.227 &   0.026 &    0.201  &   -0.042  &  0.187  & -0.043 &  0.158 & -0.043 \\
                                                                                    
 5 &   2 &  1 &   2 &   4 &    0.197  &  -0.139 &    0.228 &   0.003 &    0.207  &   -0.060  &  0.197  & -0.059 &  0.169 & -0.059 \\
                                                                                    
 5 &   2 &  0 &   3 &   6 &    0.206  &  -0.147 &    0.227 &  -0.013 &    0.212  &   -0.073  &  0.205  & -0.070 &  0.180 & -0.070 \\
\hline                                                                              
                                                                                    
 6 &   2 &  2 &   2 &   4 &    0.182  &  -0.127 &    0.213 &   0.012 &    0.192  &   -0.049  &  0.181  & -0.049 &  0.155 & -0.049 \\
                                                                                    
 6 &   2 &  1 &   3 &   6 &    0.188  &  -0.135 &    0.209 &  -0.004 &    0.194  &   -0.062  &  0.187  & -0.060 &  0.161 & -0.060 \\
                                                                                    
 6 &   2 &  0 &   4 &   8 &    0.193  &  -0.139 &    0.206 &  -0.016 &    0.195  &   -0.070  &  0.191  & -0.067 &  0.168 & -0.067 \\
\hline                                                                              
                                                                                    
 7 &   2 &  2 &   3 &   6 &    0.169  &  -0.117 &    0.193 &   0.011 &    0.176  &   -0.046  &  0.168  & -0.044 &  0.143 & -0.044 \\
                                                                                    
 7 &   2 &  1 &   4 &   8 &    0.172  &  -0.122 &    0.186 &  -0.001 &    0.174  &   -0.055  &  0.170  & -0.052 &  0.147 & -0.052 \\
                                                                                    
 7 &   2 &  0 &   5 &  10 &    0.173  &  -0.123 &    0.180 &  -0.008 &    0.173  &   -0.059  &  0.171  & -0.055 &  0.150 & -0.055 \\
\hline                                                                              
                                                                                    
 8 &   2 &  2 &   4 &   8 &    0.151  &  -0.101 &    0.178 &   0.019 &    0.156  &   -0.035  &  0.148  & -0.032 &  0.126 & -0.032 \\
                                                                                    
 8 &   2 &  1 &   5 &  10 &    0.150  &  -0.103 &    0.158 &   0.010 &    0.150  &   -0.040  &  0.147  & -0.037 &  0.126 & -0.037 \\
                                                                                    
 8 &   2 &  0 &   6 &  12 &    0.149  &  -0.102 &    0.150 &   0.006 &    0.146  &   -0.042  &  0.146  & -0.038 &  0.127 & -0.038 \\
\hline                                                                              
                                                                                    
 9 &   2 &  2 &   5 &  10 &    0.131  &  -0.080 &    0.169 &   0.032 &    0.145  &   -0.018  &  0.132  & -0.015 &  0.110 & -0.015\\
                                                                                    
 9 &   2 &  1 &   6 &  12 &    0.125  &  -0.080 &    0.143 &   0.027 &    0.127  &   -0.020  &  0.122  & -0.017 &  0.102 & -0.017 \\
\hline                                                                              
                                                                                    
 10 &  2 &  2 &   6 &  12 &   0.123  &   -0.056 &    0.166 &  -0.051 &    0.141  &    0.004  &  0.126  &  0.007 &  0.106 &  0.007 \\

\hline
\hline

\end{tabular}  

\end{table*}


\begin{figure}[tbp]
\begin{center}
\includegraphics[height=3.1367in,width=3.6in]{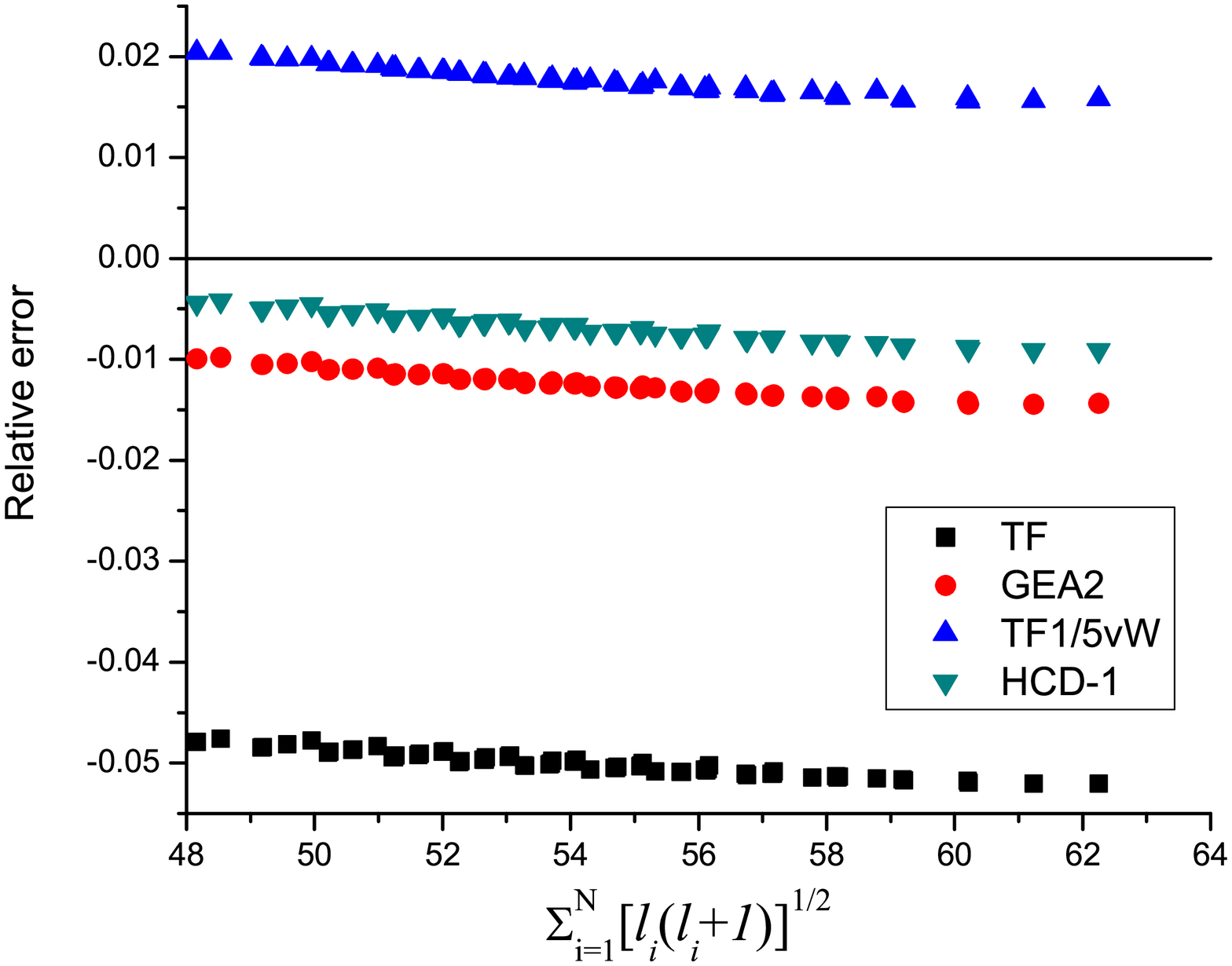}
\end{center}
\caption{ (Color online) 
Relative errors obtained with the TF, GEA2, TF$\frac{1}{5}$vW 
and the first order correction HCD functionals 
in terms of 
$\sum_{i=1}^{N} \sqrt{ l_{i}(l_{i}+1) }$, 
for all the $64$ configurations of the confined noninteracting system 
with $34$ fermions. 
Note that these results do not discriminate configurations 
with different values of $p_{4s}$, as shown in the results for 
$\sigma$ of 
Fig.(\ref{Fig:sigma_fermions})
}
\label{Fig:err_fermions}
\end{figure}

For the sake of simplicity we will only present the detailed 
results for the noninteracting systems with a number of 
fermions $N$ between $1$ and $10$. 
In Table~\ref{Table:TableES} the occupation 
numbers of each of the subshells of single-fermion orbitals 
$1s$, $2s$ and $2p$, 
and the value of $\sum_{i=1}^{N}{\sqrt{l_{i}(l_{i}+1)}}$
for each of the $N$-fermions configuration are shown. 
As commented in Ref.~(\onlinecite{2009TMBDGAJEA}), 
$\sum_{i=1}^{N}{\sqrt{l_{i}(l_{i}+1)}}$ can be intuitively 
related to an \emph{angular momentum} of the configuration, 
although it is not the expected value of the angular 
momentum operator but the sum of the eigenvalues of 
$\sqrt{L^{2}}$ 
for each independent single-fermion orbital. 
In the table we also show the relative errors $\epsilon$ 
obtained with the KEDFs and the values of 
$\sigma_{min} $ 
that give the closest $t_{S}^{L, \alpha}(\mathbf{r})$ in 
Eq.~(\ref{Eq:tslap}) for each functional.
For the TF functional, the relative errors get smaller when the 
number of fermions increases.  
Showing much smaller absolute errors
(probably due to a compensation of the TF errors by the opposite trend 
of the vW functional: its errors grow with the number of fermions%
\cite{2009TMBDGAJEA}), 
the TF$\frac{1}{5}$vW functional 
yields a similar behavior, except that the errors slightly 
increase for $N>8$.
The GEA2 and both HCD functionals present no clear trends, 
because they appear to work better for those $N$-fermion 
configurations that are close to the closed-shell.
And for all the functionals but the TF$\frac{1}{5}$vW functional 
the relative error is smaller for configurations with 
smaller values of 
$\sum_{i=1}^{N}{\sqrt{l_{i}(l_{i}+1)}}$.
Moreover, note that the errors are much smaller for the 
TF$\frac{1}{5}$vW functional than for the GEA2 one, 
an opposite trend that we found for light atoms. 

As explained in Section \ref{Sec:study}, $\sigma_{min}$ 
is the amount of approximate KED misplaced with respect to the 
distribution $t_{S}^{L, \alpha}(\mathbf{r})$ after minimizing 
with respect to the parameter $\alpha $ 
--- see Eq.~(\ref{Eq:sigma}).
The values of the quality factor of Table~\ref{Table:TableES} 
show that $\sigma_{min}$ decreases with the number of fermions 
for the TF functional.
Even the relative errors of the kinetic energy are much smaller 
than for the TF case, the TF$\frac{1}{5}$vW functional yields 
values of $\sigma$ that are bigger than those obtained for the 
TF functional, following the general results shown by the 
GGA functionals. \cite{2007DGAJEA1}
Again, the local compensation of the TF behavior and the opposite 
trend of the vW functional \cite{2009TMBDGAJEA} 
cause the result for the TF$\frac{1}{5}$vW functional.
These results for $\sigma_{min}$, and the previous mentioned 
relative errors, arise from the known properties of the vW 
functional: it is a lower bound to both the 
total KE and the local KED. 
On the other hand, the general trend for the GEA2 and both 
first-order HCD functionals indicates that the quality factor 
decreases with the number of fermions, 
with quite similar results for the three approximations. 

\begin{figure}[tbp]
\begin{center}
\includegraphics[height=3.1176in,width=3.6in]{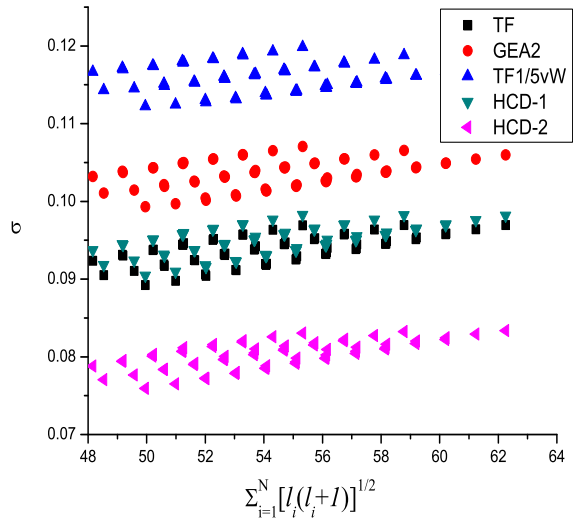}
\end{center}
\caption{ (Color online) 
Quality factor $\sigma$ 
obtained with the TF, GEA2, TF$\frac{1}{5}$vW 
and the first order correction HCD functionals 
in terms of 
$\sum_{i=1}^{N} \sqrt{ l_{i}(l_{i}+1) }$, 
for the $64$ configurations of the noninteracting $34$-fermion 
system. 
Note the different results the two HCD functionals yield.
}
\label{Fig:sigma_fermions}
\end{figure}

Focusing on the dependence on the \textsl{angular momentum} 
$\sum_{i=1}^{N}{\sqrt{l_{i}(l_{i}+1)}}$ 
of the $N$-fermion configurations we see that the local 
description of the TF functional improves for 
smaller \textsl{angular momentum} 
(i.~e. the quality factors decrease with the angular momentum). 
The TF$\frac{1}{5}$vW functional shows the opposite trend. 
On the other hand, the values of $\sigma$ for the GEA2 and 
the two HCD functionals  
decrease with the \textsl{angular momentum} when the shell 
is less than half-filled and increases when it is more than 
half-filled. 

In order to make some other comparisons on the 
dependence on the \textsl{angular momentum}, 
let us discuss the case with $N=34$ fermions, 
that has $64$ different configurations, all degenerate 
with the same energy. 
As commented before, these degenerate 
$N$-fermion systems have spherically symmetric densities, 
but due to the different occupancy of the $l$ single-fermion states 
they give different results for the kinetic energies and for 
the values of $\sigma$. 
Now we will focus on how the results obtained with each of the 
different KEDFs depend on 
$\sum_{i=1}^{N}{\sqrt{l_{i}(l_{i}+1)}}$. 
Regarding the relative errors, the HCD $r$-dependent functionals 
obviously give the same value since, once integrated, both 
yield the same kinetic energy.
Fig.~\ref{Fig:err_fermions} shows the values of the relative 
errors for the functionals for the noninteracting fermion system, 
as a function of 
$\sum_{i=1}^{N}\sqrt{l_{i}(l_{i}+1)}$. 
In this case, the HCD $r$-dependent functional gives the best 
results, better than the GEA2 ones, 
whereas the TF$\frac{1}{5}$vW functional improves the values 
obtained with the TF one, getting again positive values.
Note that for this system all the values of the relative errors 
are much smaller than for the atomic system.

Fig.~\ref{Fig:sigma_fermions} depicts $\sigma $ as a function 
of $\sum_{i=1}^{N}\sqrt{l_{i}(l_{i}+1)}$. 
In this case, the TF and HCD-1 functionals give very close results for 
$\sigma$, whereas the GEA2 functional yields worse values than the TF 
and the results of the TF$\frac{1}{5}$vW functional clearly 
go bigger.
To better understand the results, let us analyze how the fermions 
distribute among all the single-fermion orbitals.
For $N=34$, the fermions occupy the lowest-energy orbitals, 
with six fermions in the fourth shell. 
Now we put our attention on the configurations that have a fixed value 
of $\sum_{i=1}^{N}{\sqrt{l_{i}(l_{i}+1)}}$, 
i.~e. those results with the same value for the abscissas in the 
figure. 
Using the arguments presented in Ref.~(\onlinecite{2009TMBDGAJEA}), 
for a fixed value of $\sum_{i=1}^{N}{\sqrt{l_{i}(l_{i}+1)}}$ 
the configurations with $n_{4s}=2$ have more high-$l_{i}$ single-fermion 
orbitals occupied than the configurations with $n_{4s}=1$; 
and when $n_{4s}=0$ we will have our six fermions 
occupying orbitals with the lowest-$l_{i}$ possible values.
Fig.~\ref{Fig:sigma_fermions} then shows 
that the values of $\sigma$ for the distinct configurations 
are clearly ordered in terms of the different values of 
$p_{4s}$. 
This is not the case for the relative error of the energies, 
as can be seen in Fig.~\ref{Fig:err_fermions}, where 
no differences are observed. 
This effect was also previously studied in 
Ref.~(\onlinecite{2009TMBDGAJEA}).

\section{Conclusions}

The first-order correction HCD functionals 
improve the semilocal 
functionals (GEA2 and TF$\frac{1}{5}$vW) we have used as 
benchmarks in the test systems we have studied. 
The relative errors are quite similar to the GEA2 functional 
for atoms, whereas they are slightly smaller for the 
noninteracting confined fermion system. 
We get accurate total kinetic energies, 
but when the local KEDs behavior is measured through $\sigma$, 
only the HCD-2 functional clearly gives better values than the 
TF functional, 
whereas the HCD-1 functional gets quite close results to those 
of the TF one. 
These results have revealed the differences in the local behavior 
of both HCD functionals. 
On the other hand, Figs.~\ref{Fig:sigma_atoms} and 
\ref{Fig:sigma_fermions} 
show that the values of $\sigma$ are clearly smaller for the 
noninteracting fermion system than for the atomic system, 
reflecting the importance of the interaction on the KED.

As was also found in the previous paper,\cite{2009TMBDGAJEA} 
we have shown that for a specific number of fermions 
the results for the KEDs generally are different 
for each of the degenerate states of the 
noninteracting $N$-fermion system 
with spherical symmetric densities. 
Depicting their values of $\sigma$ versus the 
values of  $\sum_{i=1}^{N}\sqrt{l_{i}(l_{i}+1)}$ 
of each specific configuration 
some trends appear: 
for a given $\sum_{i=1}^{N}{\sqrt{l_{i}(l_{i}+1)}}$, 
those configurations whose occupied orbitals have 
the lowest values of $l_{i}$ compatible this value of  
$\sum_{i=1}^{N}{\sqrt{l_{i}(l_{i}+1)}}$ 
yield the smaller values for $\sigma$.
As a consequence, the KEDFs discussed in the paper give 
a better description of the KED when the 
lower-$l_{i}$ single-fermion orbitals have the 
maximum occupancy.
These KEDFs give a better description of the KED when 
the lower-$l_{i}$ single-fermion orbitals have the 
maximum occupancy.

An open question is to understand the reasons why one of 
these first-order correction HCD functionals gives better 
local behavior than any other KEDF in the literature. 


\begin{acknowledgments}
Prof.~R.~G.~Parr (University of North Carolina at Chapel Hill, USA) 
is grateful acknowledged by suggesting the study of the First-Order Gradient 
Correction to the Thomas-Fermi Kinetic Energy Functional that we present in 
this paper.
This work has been partially supported by a grant of the 
Ministerio de Educaci\'{o}n y Ciencia of Spain 
(reference FIS2007-65702-C02-02).
\end{acknowledgments}



\begin{thebibliography}{0}
\expandafter\ifx\csname natexlab\endcsname\relax\def\natexlab#1{#1}\fi
\expandafter\ifx\csname bibnamefont\endcsname\relax
  \def\bibnamefont#1{#1}\fi
\expandafter\ifx\csname bibfnamefont\endcsname\relax
  \def\bibfnamefont#1{#1}\fi
\expandafter\ifx\csname citenamefont\endcsname\relax
  \def\citenamefont#1{#1}\fi
\expandafter\ifx\csname url\endcsname\relax
  \def\url#1{\texttt{#1}}\fi
\expandafter\ifx\csname urlprefix\endcsname\relax\def\urlprefix{URL }\fi
\providecommand{\bibinfo}[2]{#2}
\providecommand{\eprint}[2][]{\url{#2}}

\end{thebibliography}


\begin{thebibliography}{33}
\expandafter\ifx\csname natexlab\endcsname\relax\def\natexlab#1{#1}\fi
\expandafter\ifx\csname bibnamefont\endcsname\relax
  \def\bibnamefont#1{#1}\fi
\expandafter\ifx\csname bibfnamefont\endcsname\relax
  \def\bibfnamefont#1{#1}\fi
\expandafter\ifx\csname citenamefont\endcsname\relax
  \def\citenamefont#1{#1}\fi
\expandafter\ifx\csname url\endcsname\relax
  \def\url#1{\texttt{#1}}\fi
\expandafter\ifx\csname urlprefix\endcsname\relax\def\urlprefix{URL }\fi
\providecommand{\bibinfo}[2]{#2}
\providecommand{\eprint}[2][]{\url{#2}}

\bibitem[{\citenamefont{Hohenberg and Kohn}(1964)}]{1964HK}
\bibinfo{author}{\bibfnamefont{P.}~\bibnamefont{Hohenberg}} \bibnamefont{and}
  \bibinfo{author}{\bibfnamefont{W.}~\bibnamefont{Kohn}},
  \bibinfo{journal}{Phys. Rev. B} \textbf{\bibinfo{volume}{136}},
  \bibinfo{pages}{864} (\bibinfo{year}{1964}).

\bibitem[{\citenamefont{Kohn and Sham}(1965)}]{1965KS}
\bibinfo{author}{\bibfnamefont{W.}~\bibnamefont{Kohn}} \bibnamefont{and}
  \bibinfo{author}{\bibfnamefont{L.~J.} \bibnamefont{Sham}},
  \bibinfo{journal}{Phys. Rev. A} \textbf{\bibinfo{volume}{140}},
  \bibinfo{pages}{1133} (\bibinfo{year}{1965}).

\bibitem[{\citenamefont{Garc\'{\i}a-Aldea and
  Alvarellos}(2007{\natexlab{a}})}]{2007DGAJEA1}
\bibinfo{author}{\bibfnamefont{D.}~\bibnamefont{Garc\'{\i}a-Aldea}}
  \bibnamefont{and} \bibinfo{author}{\bibfnamefont{J.~E.}
  \bibnamefont{Alvarellos}}, \bibinfo{journal}{J. Chem. Phys.}
  \textbf{\bibinfo{volume}{127}}, \bibinfo{pages}{144109}
  (\bibinfo{year}{2007}{\natexlab{a}}).

\bibitem[{\citenamefont{Haq et~al.}(1984)\citenamefont{Haq, Chattaraj, and
  Deb}}]{1984HCD}
\bibinfo{author}{\bibfnamefont{S.}~\bibnamefont{Haq}},
  \bibinfo{author}{\bibfnamefont{P.~K.} \bibnamefont{Chattaraj}},
  \bibnamefont{and} \bibinfo{author}{\bibfnamefont{B.~M.} \bibnamefont{Deb}},
  \bibinfo{journal}{Chem. Phys. Lett.} \textbf{\bibinfo{volume}{81}},
  \bibinfo{pages}{8028} (\bibinfo{year}{1984}).

\bibitem[{\citenamefont{Deb and Chattaraj}(1988)}]{1988DC}
\bibinfo{author}{\bibfnamefont{B.~M.} \bibnamefont{Deb}} \bibnamefont{and}
  \bibinfo{author}{\bibfnamefont{P.~K.} \bibnamefont{Chattaraj}},
  \bibinfo{journal}{Phys. Rev. A} \textbf{\bibinfo{volume}{37}},
  \bibinfo{pages}{4030} (\bibinfo{year}{1988}).

\bibitem[{\citenamefont{Chattaraj}(1990)}]{1990Chattaraj}
\bibinfo{author}{\bibfnamefont{P.~K.} \bibnamefont{Chattaraj}},
  \bibinfo{journal}{Phys Rev A.} \textbf{\bibinfo{volume}{41}},
  \bibinfo{pages}{6505} (\bibinfo{year}{1990}).

\bibitem[{\citenamefont{Deb and Chattaraj}(1992)}]{1992DC}
\bibinfo{author}{\bibfnamefont{B.~M.} \bibnamefont{Deb}} \bibnamefont{and}
  \bibinfo{author}{\bibfnamefont{P.~K.} \bibnamefont{Chattaraj}},
  \bibinfo{journal}{Phys. Rev. A} \textbf{\bibinfo{volume}{45}},
  \bibinfo{pages}{1412} (\bibinfo{year}{1992}).

\bibitem[{\citenamefont{Zhou et~al.}(1992)\citenamefont{Zhou, Chattaraj, Parr,
  and Lee}}]{1992ZCPL}
\bibinfo{author}{\bibfnamefont{Z.}~\bibnamefont{Zhou}},
  \bibinfo{author}{\bibfnamefont{P.~K.} \bibnamefont{Chattaraj}},
  \bibinfo{author}{\bibfnamefont{R.~G.} \bibnamefont{Parr}}, \bibnamefont{and}
  \bibinfo{author}{\bibfnamefont{C.}~\bibnamefont{Lee}},
  \bibinfo{journal}{Theor. Chim. Acta} \textbf{\bibinfo{volume}{84}},
  \bibinfo{pages}{237} (\bibinfo{year}{1992}).

\bibitem[{\citenamefont{Garc\'{\i}a-Aldea and
  Alvarellos}(2007{\natexlab{b}})}]{2007DGAJEA2}
\bibinfo{author}{\bibfnamefont{D.}~\bibnamefont{Garc\'{\i}a-Aldea}}
  \bibnamefont{and} \bibinfo{author}{\bibfnamefont{J.~E.}
  \bibnamefont{Alvarellos}}, \bibinfo{journal}{Phys. Rev. A}
  \textbf{\bibinfo{volume}{76}}, \bibinfo{pages}{052504}
  (\bibinfo{year}{2007}{\natexlab{b}}).

\bibitem[{\citenamefont{Garc\'{\i}a-Aldea and
  Alvarellos}(2008{\natexlab{a}})}]{2008DGAJEA1}
\bibinfo{author}{\bibfnamefont{D.}~\bibnamefont{Garc\'{\i}a-Aldea}}
  \bibnamefont{and} \bibinfo{author}{\bibfnamefont{J.~E.}
  \bibnamefont{Alvarellos}}, \bibinfo{journal}{Phys. Rev. A}
  \textbf{\bibinfo{volume}{77}}, \bibinfo{pages}{022502}
  (\bibinfo{year}{2008}{\natexlab{a}}).

\bibitem[{\citenamefont{Garc\'{\i}a-Aldea and
  Alvarellos}(2008{\natexlab{b}})}]{2008DGAJEA2}
\bibinfo{author}{\bibfnamefont{D.}~\bibnamefont{Garc\'{\i}a-Aldea}}
  \bibnamefont{and} \bibinfo{author}{\bibfnamefont{J.~E.}
  \bibnamefont{Alvarellos}}, \bibinfo{journal}{J. Chem. Phys.}
  \textbf{\bibinfo{volume}{129}}, \bibinfo{pages}{074103}
  (\bibinfo{year}{2008}{\natexlab{b}}).

\bibitem[{\citenamefont{Mart\'{\i}n-Blas
  et~al.}(2009)\citenamefont{Mart\'{\i}n-Blas, Garc\'{\i}a-Aldea, and
  Alvarellos}}]{2009TMBDGAJEA}
\bibinfo{author}{\bibfnamefont{T.}~\bibnamefont{Mart\'{\i}n-Blas}},
  \bibinfo{author}{\bibfnamefont{D.}~\bibnamefont{Garc\'{\i}a-Aldea}},
  \bibnamefont{and} \bibinfo{author}{\bibfnamefont{J.~E.}
  \bibnamefont{Alvarellos}}, \bibinfo{journal}{J. Chem. Phys.}
  \textbf{\bibinfo{volume}{130}}, \bibinfo{pages}{034101}
  (\bibinfo{year}{2009}).

\bibitem[{\citenamefont{Thomas}(1927)}]{1927Th}
\bibinfo{author}{\bibfnamefont{L.~H.} \bibnamefont{Thomas}},
  \bibinfo{journal}{Proc. Camb. Phil. Soc.} \textbf{\bibinfo{volume}{23}},
  \bibinfo{pages}{542} (\bibinfo{year}{1927}).

\bibitem[{\citenamefont{Fermi}(1927)}]{1927Fer}
\bibinfo{author}{\bibfnamefont{E.}~\bibnamefont{Fermi}},
  \bibinfo{journal}{Rend. Accad. Lincei} \textbf{\bibinfo{volume}{6}},
  \bibinfo{pages}{602} (\bibinfo{year}{1927}).

\bibitem[{\citenamefont{Weizsacker}(1935)}]{1935Weiz}
\bibinfo{author}{\bibfnamefont{C.~F.~V.} \bibnamefont{Weizsacker}},
  \bibinfo{journal}{Z. Physik} \textbf{\bibinfo{volume}{96}},
  \bibinfo{pages}{431} (\bibinfo{year}{1935}).

\bibitem[{\citenamefont{Kirzhnits}(1957)}]{1957Kirzhnits}
\bibinfo{author}{\bibfnamefont{D.~A.} \bibnamefont{Kirzhnits}},
  \bibinfo{journal}{Sov. Phys.- JETP} \textbf{\bibinfo{volume}{5}},
  \bibinfo{pages}{64} (\bibinfo{year}{1957}).

\bibitem[{\citenamefont{Yonei and Tomishima}(1965)}]{1965YoneiTomish}
\bibinfo{author}{\bibfnamefont{K.}~\bibnamefont{Yonei}} \bibnamefont{and}
  \bibinfo{author}{\bibfnamefont{Y.}~\bibnamefont{Tomishima}},
  \bibinfo{journal}{J. Phys. Soc. Jpn.} \textbf{\bibinfo{volume}{20}},
  \bibinfo{pages}{1051} (\bibinfo{year}{1965}).

\bibitem[{\citenamefont{Tomishima and Yonei}(1966)}]{1966TomishYonei}
\bibinfo{author}{\bibfnamefont{Y.}~\bibnamefont{Tomishima}} \bibnamefont{and}
  \bibinfo{author}{\bibfnamefont{K.}~\bibnamefont{Yonei}}, \bibinfo{journal}{J.
  Phys. Soc. Jpn.} \textbf{\bibinfo{volume}{21}}, \bibinfo{pages}{142}
  (\bibinfo{year}{1966}).

\bibitem[{\citenamefont{Gadre and Pathak}(1982)}]{1982GadrePathak}
\bibinfo{author}{\bibfnamefont{S.~R.} \bibnamefont{Gadre}} \bibnamefont{and}
  \bibinfo{author}{\bibfnamefont{R.~K.} \bibnamefont{Pathak}},
  \bibinfo{journal}{Phys. Rev. A} \textbf{\bibinfo{volume}{25}},
  \bibinfo{pages}{668} (\bibinfo{year}{1982}).

\bibitem[{\citenamefont{Fisher}(1925)}]{1925Fisher}
\bibinfo{author}{\bibfnamefont{R.~A.} \bibnamefont{Fisher}},
  \bibinfo{journal}{Proc. Cambridge Philos. Soc.}
  \textbf{\bibinfo{volume}{22}}, \bibinfo{pages}{700} (\bibinfo{year}{1925}).

\bibitem[{\citenamefont{Sears et~al.}(1980)\citenamefont{Sears, Parr, and
  Dinur}}]{1980SearsParrDinur}
\bibinfo{author}{\bibfnamefont{S.~B.} \bibnamefont{Sears}},
  \bibinfo{author}{\bibfnamefont{R.~G.} \bibnamefont{Parr}}, \bibnamefont{and}
  \bibinfo{author}{\bibfnamefont{V.}~\bibnamefont{Dinur}},
  \bibinfo{journal}{Isr. J. Chem.} \textbf{\bibinfo{volume}{19}},
  \bibinfo{pages}{165} (\bibinfo{year}{1980}).

\bibitem[{\citenamefont{Liu}(2007)}]{2007L}
\bibinfo{author}{\bibfnamefont{S.}~\bibnamefont{Liu}}, \bibinfo{journal}{J.
  Chem. Phys.} \textbf{\bibinfo{volume}{126}}, \bibinfo{pages}{191107}
  (\bibinfo{year}{2007}).

\bibitem[{\citenamefont{Romera and Dehesa}(1994)}]{1994RomeraDehesa}
\bibinfo{author}{\bibfnamefont{E.}~\bibnamefont{Romera}} \bibnamefont{and}
  \bibinfo{author}{\bibfnamefont{J.~S.} \bibnamefont{Dehesa}},
  \bibinfo{journal}{Phys. Rev. A} \textbf{\bibinfo{volume}{50}},
  \bibinfo{pages}{256} (\bibinfo{year}{1994}).

\bibitem[{\citenamefont{Yang et~al.}(1996)\citenamefont{Yang, Liu, and
  Wang}}]{1996YLW}
\bibinfo{author}{\bibfnamefont{Z.-Z.} \bibnamefont{Yang}},
  \bibinfo{author}{\bibfnamefont{S.}~\bibnamefont{Liu}}, \bibnamefont{and}
  \bibinfo{author}{\bibfnamefont{Y.~A.} \bibnamefont{Wang}},
  \bibinfo{journal}{Chem. Phys. Lett.} \textbf{\bibinfo{volume}{258}},
  \bibinfo{pages}{30} (\bibinfo{year}{1996}).

\bibitem[{\citenamefont{Atkins}(1997)}]{b1997Atkins}
\bibinfo{author}{\bibfnamefont{P.~W.} \bibnamefont{Atkins}},
  \emph{\bibinfo{title}{Molecular Quantum Mechanics}}
  (\bibinfo{publisher}{Oxford University Press, 3rd. ed.},
  \bibinfo{address}{Oxford}, \bibinfo{year}{1997}).

\bibitem[{\citenamefont{Clementi and Raimondi}(1963)}]{1963CR}
\bibinfo{author}{\bibfnamefont{E.}~\bibnamefont{Clementi}} \bibnamefont{and}
  \bibinfo{author}{\bibfnamefont{D.~L.} \bibnamefont{Raimondi}},
  \bibinfo{journal}{J. Chem. Phys.} \textbf{\bibinfo{volume}{38}},
  \bibinfo{pages}{2686} (\bibinfo{year}{1963}).

\bibitem[{\citenamefont{Lieb}(1983)}]{1983Lieb}
\bibinfo{author}{\bibfnamefont{E.~H.} \bibnamefont{Lieb}},
  \bibinfo{journal}{Int. J. Quantum Chem.} \textbf{\bibinfo{volume}{24}},
  \bibinfo{pages}{243} (\bibinfo{year}{1983}).

\bibitem[{\citenamefont{Levy}(1982)}]{1982Levy}
\bibinfo{author}{\bibfnamefont{M.}~\bibnamefont{Levy}}, \bibinfo{journal}{Phys.
  Rev. A} \textbf{\bibinfo{volume}{26}}, \bibinfo{pages}{1200}
  (\bibinfo{year}{1982}).

\bibitem[{\citenamefont{Englisch and Englisch}(1983)}]{1983EE}
\bibinfo{author}{\bibfnamefont{H.}~\bibnamefont{Englisch}} \bibnamefont{and}
  \bibinfo{author}{\bibfnamefont{R.}~\bibnamefont{Englisch}},
  \bibinfo{journal}{Physica A} \textbf{\bibinfo{volume}{121}},
  \bibinfo{pages}{253} (\bibinfo{year}{1983}).

\bibitem[{\citenamefont{Ulrich and Kohn}(2000)}]{2000UK}
\bibinfo{author}{\bibfnamefont{C.~A.} \bibnamefont{Ulrich}} \bibnamefont{and}
  \bibinfo{author}{\bibfnamefont{W.}~\bibnamefont{Kohn}},
  \bibinfo{journal}{Phys. Rev. Lett.} \textbf{\bibinfo{volume}{87}},
  \bibinfo{pages}{093001} (\bibinfo{year}{2000}).

\bibitem[{\citenamefont{Nagy}(1998)}]{1998Nagy}
\bibinfo{author}{\bibfnamefont{{\'{A}}.}~\bibnamefont{Nagy}},
  \bibinfo{journal}{Phys. Rev. A} \textbf{\bibinfo{volume}{57}},
  \bibinfo{pages}{1672} (\bibinfo{year}{1998}).

\bibitem[{\citenamefont{Yang et~al.}(2000)\citenamefont{Yang, Zhang, and
  Ayers}}]{2000YZA}
\bibinfo{author}{\bibfnamefont{W.}~\bibnamefont{Yang}},
  \bibinfo{author}{\bibfnamefont{Y.}~\bibnamefont{Zhang}}, \bibnamefont{and}
  \bibinfo{author}{\bibfnamefont{P.~W.} \bibnamefont{Ayers}},
  \bibinfo{journal}{Phys. Rev. Lett.} \textbf{\bibinfo{volume}{84}},
  \bibinfo{pages}{5172} (\bibinfo{year}{2000}).

\bibitem[{\citenamefont{Nagy et~al.}(2005)\citenamefont{Nagy, Liu, and
  Bartolloti}}]{2005NLB}
\bibinfo{author}{\bibfnamefont{{\'{A}}.}~\bibnamefont{Nagy}},
  \bibinfo{author}{\bibfnamefont{S.}~\bibnamefont{Liu}}, \bibnamefont{and}
  \bibinfo{author}{\bibfnamefont{L.}~\bibnamefont{Bartolloti}},
  \bibinfo{journal}{J. Chem. Phys.} \textbf{\bibinfo{volume}{122}},
  \bibinfo{pages}{134107} (\bibinfo{year}{2005}).

\end{thebibliography}


\newpage

\end{document}